**Full Title:  Why does women's fertility end in mid-life?  Grandmothering and age at last birth.**

Short title:  Grandmothering and age at last birth

Peter S. Kim[1*], John S. McQueen[1], Kristen Hawkes[2]

[1] School of Mathematics and Statistics, University of Sydney, NSW 2006, Australia

[2] Department of Anthropology, University of Utah, Salt Lake City, UT 84112, USA

* Corresponding author

Email: peter.kim@sydney.edu.au

Keywords:
Human evolution
Life history
Agent-based modeling
Grandmothering
Menopause
Human longevity

**Abstract**

Great apes, the other living members of our hominid family, become decrepit before the age of forty and rarely outlive their fertile years. In contrast, women – even in high mortality hunter-gatherer populations – usually remain healthy and productive well beyond menopause. The grandmother hypothesis aims to account for the evolution of this distinctive feature of human life history. Our previous mathematical simulations of that hypothesis fixed the end of female fertility at the age of 45, based on the similarities among living hominids, and then modeled the evolution of human-like longevity from an ancestral state, like that of the great apes, due only to grandmother effects. A major modification here allows the age female fertility ends to vary as well, directly addressing a version of the question, influentially posed by GC Williams six decades ago: Why isn't menopause later in humans? Our model is an agent-based model (ABM) that accounts for the coevolution of both expected adult lifespan and end of female fertility as selection maximizes reproductive value. We find that grandmother effects not only drive the population from an equilibrium representing a great ape-like longevity to a new human-like longevity, they also maintain the observed termination of women's fertility before the age of 50.

**Introduction**

Evolutionary questions raised by a distinctive post-fertile life stage in women (1, 2) have long drawn the attention of life-history theoreticians (3). We develop a mathematical model to engage the question posed sixty years ago by GC Williams (4): Why do women have post-fertile lifespans? This question has been restated and specified in two common ways: (1) Why did human longevity evolve to extend beyond the end of female fertility, and (2) Why do women stop giving birth early? The first supposes the end of female fertility to be an ancestral trait humans share with the other members of our hominid radiation where female fertility also ends in the forties, while our greater longevity is derived. The second question supposes a long-lived ancestor in which female fertility continued longer, then evolved to end at earlier ages.

In his 1957 paper, Williams deduced that, "There should be little or no post-reproductive period in the normal life-cycle of any species" for which conditions in his theory applied (4, p. 407). He went on to consider the apparent falsification presented by our own species, first pointing out that "any individual caring for dependent offspring is acting in a way that promotes the survival of his own genes and is properly considered a part of the breeding population," and so, "at some time during human evolution it may have become advantageous for a woman of forty-five or fifty to stop dividing her declining faculties between the care of extant offspring and the production of new ones" (4, p. 407).

In 1957, menopause was thought uniquely human. Evidence now shows that other primates also reach menopause, if they live long enough – a pattern best documented in macaques (5). In human populations practicing natural fertility, average ages at last birth and then at cessation of menstrual cycling are about ten years apart (6-8). While as in other mammals (9), human males continue to produce new gametes throughout life, mammalian females establish a fixed stock of oocytes in early development that depletes continuously thereafter (10). Women's fertility ends when that oocyte stock falls to thresholds insufficient to maintain menstrual cycles (11, 12).

Phylogenetic evidence also unavailable to Williams in 1957 now confirms that great apes belong in our hominid family (13, 14), and we all share similar oldest ages of parturition (15, 16). This is grounds for inferring that humans don't "stop early." Instead, we retain the ancestral condition. Human longevity is, however, exceptional (2, 17-20). These lines of evidence favor framing the question about human evolution the first way: Why did human longevity evolve to extend beyond the end of female fertility?

Yet Williams's stopping early question can be revised to ask: If age at menopause hardly changed in human evolution, then why not? Although humans are long-lived animals, other long-lived mammals extend female fertility to ages well beyond those reached in our lineage. Elephants, for example, continue giving birth into their sixties (21, 22), and Antarctic fin whales have been found pregnant into their eighties (23). As noted by Hawkes (18), this wider variation shows it is not mammalian physiology that constrains female fertility to end by 45. Instead, the empirical pattern suggests an underlying evolutionary tradeoff.

The increased-longevity formulation of the grandmother hypothesis (3, 24, 25) was initially stimulated by observations of the economic productivity of postmenopausal Hadza hunter-gatherer women (26) and the importance of their contribution especially when mothers bore new infants (27). The hypothesis takes advantage of Charnov's (28, 29) insights about tradeoffs that characterize and help explain life history variation across female mammals. In Charnov's mammal model, adult mortality or expected adult lifespan determines optimal age at first birth and age at juvenile independence. These relationships are consistent with increased longevity depending on greater allocation to maintenance and repair at the expense of current reproduction (4, 30-32). Charnov's model assumes that lower adult mortality favors delaying maturity to continue to grow longer and reach a larger size before reproducing at a cost of increased generation time and slower offspring production (33).

The grandmother hypothesis uses this framework to explain why in ecological circumstances where key foods cannot be effectively obtained and processed by young juveniles at rates sufficient to make them independent feeders, provisioning of dependents by post-fertile females would raise the rate of offspring production by shortening birth intervals during a younger female's fertile years. As longer-lived females could help more, they would leave more descendants, increasing expected adult lifespans in subsequent generations, lowering the cost of delaying maturity (24).

In previous work, we developed an agent-based model in which we fixed the end of child bearing at 45 years to see if the evolution of human longevity could be propelled by grandmothers' provisioning alone (34, 35). Subsequently, Chan et al. (36) formulated a deterministic partial differential equation (PDE) model of this system. In this paper, we modify the agent-based model of Kim et al. (35) so that the age female fertility ends and expected adult lifespan are both heritable traits that can mutate. Unlike our previous models, we also track matrilineal lineages to simulate matrilineal grandmothering and matrilateral aunting.

**Model**

*General overview*

Our model is formulated as a probabilistic agent-based model (ABM) written in the programming language Go (https://golang.org) (37). It is updated according to the Gillespie algorithm, so that the algorithm randomly determines the next event, e.g., next birth, next death, or next life-stage transition, and variable time step to the next event (38). It has many of the features of our model in (35), but the key modeling differences are as follows: (1) the age $M$ when a female's fertility ends is a heritable trait that can vary by mutation, (2) the two heritable traits of an individual's expected adult lifespan, $L$, and a female's end of fertility, $M$, are inherited as two alleles from the individual's mother and father, (3) there is a tradeoff between the age a female's fertility terminates and her average conception rate during her fertile years, (4) post-fertile females only grandmother matrilineal descendants, rather than being randomly paired with any dependent, and (5) post-fertile females can care for more than one dependent at a time.

The additions to the model described above cause shifts in the locations of long-term equilibria and the conditions under which equilibria are stable, so we also add a slight change in the scaling of the life history parameters from Kim et al. (35) to bring the equilibria near great ape-like and human-like values.

*Life stages*

Each individual progresses through several life stages: unweaned dependent, weaned dependent, juvenile, fertile adult and post-fertile adult (females only), eligible adult (males only), and geriatrically infirm adult. Dependent offspring are those still dependent on food subsidies from an adult female for survival. Unweaned dependents still require their mothers for nursing, but weaned dependents can be supported by either their mothers or grandmothers. Juveniles no longer require subsidies, but are not yet of reproductive age. Fertile adults (females) are sexually mature and available to reproduce. Post-fertile females have passed the end of their fertility, but can still contribute to supporting dependents. By eligible adults (males), we mean males who are able to compete with other males for paternities. Geriatrically infirm males and females are too infirm to contribute any longer to the population.

The first transitions from unweaned to weaned dependent and from dependent to juvenile occur at the ages of weaning, $\tau_0(L)$, and juvenile independence, $\tau_1(L)$, which are functions of the individual's expected adult lifespan, $L$. (See Figure 1 for a diagram of the dependent ages.) For simplicity, we assume mortality rates are constant, so that each individual has a lifetime mortality rate of $1/L$. In addition, every individual is subject to a density-dependent death rate that affects everyone equally. Specifically, whenever the population surpasses a carrying capacity, $K$, the algorithm randomly selects with equal probability an individual who has reached independence and removes him or her from the system. If the individual is a female with a dependent, the dependent is also removed.

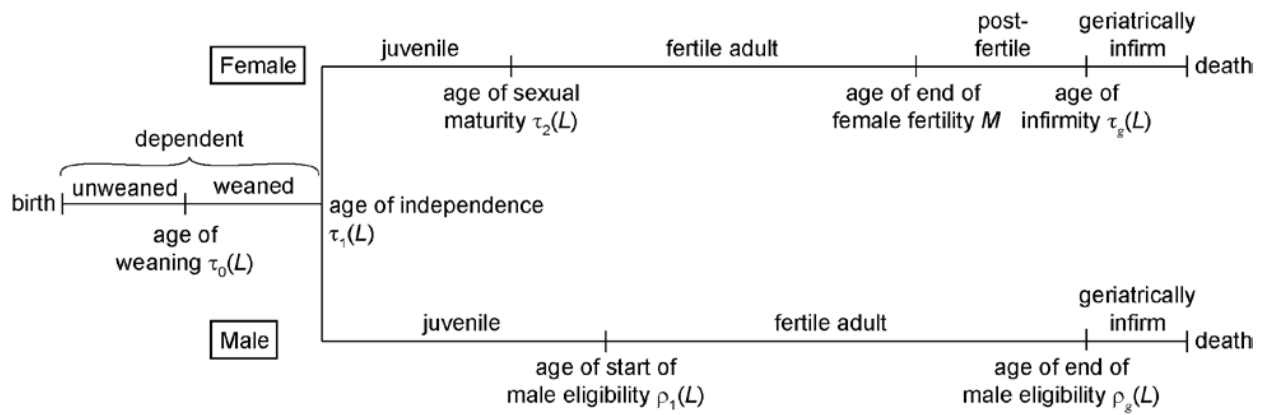

**Figure 1.** Diagram of life stages for females and males (timeline not drawn to scale). See Table 1 for definitions of $\tau_0(L)$, $\tau_1(L)$, $\tau_2(L)$, $\tau_3(L)$, $\tau_g(L)$, $\rho_1(L)$, $\rho_g(L)$, and $M$.

*Mating*

A female is fertile between the age of sexual maturity, $\tau_2(L)$, and the age $M$ when her fertility ends. A female's $M$ is a heritable trait that is calculated from alleles inherited from both parents (discussed in the next subsection). A post-fertile female, aged $M$ to her age of infirmity $\tau_g(L)$, is eligible to grandmother, i.e., subsidize a weaned dependent. A male is eligible to compete for paternities between the ages $\rho_1(L)$ and $\rho_g(L)$. Setting upper age limits $\tau_g(L)$ and $\rho_g(L)$ counteracts the possibility of grandmothering or fathering at unrealistically high ages that may result from our assumption of a constant mortality rate (see Figure 1).

Only fertile females without dependents (weaned or unweaned) can conceive. For simplicity, we assume a female conceives at a constant rate $c(M)$, which is a function of her $M$, throughout her fertile ages. When a female is available to conceive, all eligible males compete for the paternity.

Following Williams' (4) deduction that selection for reduced senescence should decrease youthful vigor, we assume that the cost of higher $L$ in males is that they are less successful at competing for paternities. To model the male fertility-longevity tradeoff, we assign each male a weighting factor $\alpha(L)$, which is a decreasing function of the male's $L$. Then, a particular male's probability of winning a paternity is $\alpha(L)\big/\left(\sum_i \alpha(L_i)\right)$ where $L$ is the male's expected adult lifespan and the summation is taken over all eligible males at the current time.

Each individual possesses two alleles $m_1$ and $m_2$, pertaining to the end of female fertility, and two alleles $l_1$ and $l_2$, pertaining to expected adult lifespan. For each heritable trait, offspring inherit one randomly chosen allele from the mother and another randomly chosen allele from the father. The offspring's realized end of female fertility is determined by the geometric mean of its two alleles $m_1$ and $m_2$, i.e., $M = \sqrt{m_1 m_2}$. Similarly, a female offspring's expected adult lifespan is determined by $L = \sqrt{l_1 l_2}$ . Note that both females and males possess alleles for the end of female fertility, but in males, the trait remains latent and does not affect their life histories.

During reproduction, each inherited allele can mutate independently with probability $p$ and result in a shift by a random factor following a lognormal distribution with mean 0 and standard deviation $\sigma$. We assume that the pair $p$ and $\sigma$ are the same for alleles pertaining both to $L$ and $M$.

By using a geometric mean and lognormal distribution, we are applying a logarithmic scale to the traits, meaning that we measure mutations in terms of relative, rather than linear, changes.

For example, we would consider shifts in $L$ from 20 to 21 and from 40 to 42 both as 5% changes up, rather than as 1 and 2 year linear shifts.  The use of a lognormal rather than a normal distribution (or any other distribution with low variance) makes little difference in the stability and location of long-term equilibria (data not shown), and we opt to use a lognormal distribution as a straightforward way to guarantee no mutations to values below 0.

*Grandmothering*

A post-fertile female may allocate care to related, weaned dependents in two ways: (1) through food subsidies, or (2) through reduction in mortality.  If a dependent is not fully subsidized at all times by some adult (mother, grandmother, or aunt), the dependent dies.

Grandmothers have limited resources that need to be divided among grandchildren. The resources required by each child depend on the age of the child and other model variables. To model the cost to subsidize a weaned dependent, we assume each adult has a total subsidizing capacity of 1, and every weaned dependent has an age-dependent subsidizing cost of

$$C_{max} \cdot \frac{\tau_1(L) - a}{\tau_1(L) - \tau_0(L)}$$

where $a$ is the current age of the dependent and $C_{max} \leq 1$ is the cost of the subsidy when the dependent is just weaned. (See Figure 2 for a plot of the cost of support versus a dependent's age.) In addition, we assume that each grandmother can only subsidize a maximum number of $D_{max}$ dependents at once. Thus, each grandmother can subsidize up to $D_{max}$ dependents as long as their total subsidizing cost remains less than her capacity of 1.

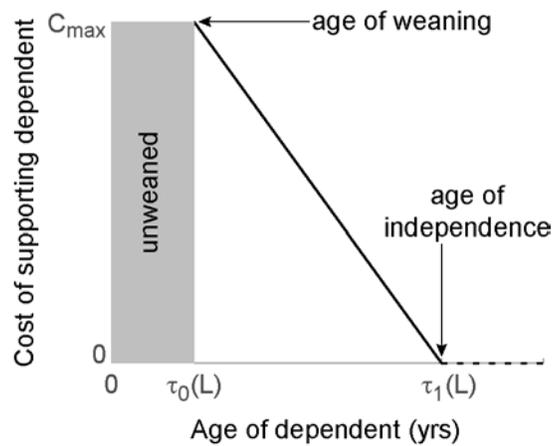

**Figure 2.** Cost of supporting a dependent as a function of the dependent's age. Between ages 0 and $\tau_0(L)$, the dependent is unweaned and cannot be sustained by a grandmother. Between ages $\tau_0(L)$ and $\tau_1(L)$, the cost of supporting the dependent drops linearly from $C_{max}$ to 0, at which point the offspring becomes independent.

In selecting available dependents, grandmothers will subsidize dependent matrilineal grandchildren first and choose from the eldest to the youngest. If there are none available, they will support dependent matrilateral nieces and nephews, i.e., sister's offspring, from the eldest to the youngest.

Beyond providing feeding subsidies, the presence of a living grandmother also results in reduced mortality of grandchildren (24, 27, 39). We model this effect by assuming eligible grandmothers reduce the mortality of dependent grandchildren. The benefit is modeled as a reduction of dependents' mortalities by a factor $\beta \in [0,1]$, which we call the grandmother benefit. Higher $\beta$ corresponds to greater benefit, and if $\beta = 1$, a grandmother's dependents are immortal while she is eligible. More generally, if a dependent has an eligible grandmother, its mortality rate is $(1-\beta)/L$, instead of $1/L$. We assume matrilineal grandmothers confer this benefit, but aunts do not, and for simplicity, we assume grandmothers provide this mortality reduction to all dependent grandchildren, not only those they are fully subsidizing. This kind of benefit, distinct from direct subsidies, might capture the increase in reproductive success observed in some nonhuman primate females whose own mothers are still alive (40, 41).

Distinguishing two forms of grandmother help, (1) full food subsidies, and (2) mortality reduction, allows us to investigate the impacts of these differing benefits. (See Table 1 for a summary of parameters.)

**Table 1.** Parameters of the agent-based model.

| Parameter | Description | Estimate |
|---|---|---|
| $L$ | Expected adult lifespan (yr) | Variable heritable trait |
| $M$ | Age female fertility ends (yr) | Variable heritable trait |
| $\tau_0(L)$ | Age of weaning (yr) | $\min\{L/8, 3\}$ |
| $\tau_1(L)$ | Age of independence (yr) | $L/5.2$ |
| $\tau_2(L)$ | Age of female sexual maturity (yr) | $L/2$ |
| $\tau_3(L, M)$ | Age female fertility ends (yr) | $\min\{M, \tau_g(L)\}$ |
| $\tau_g(L)$ | Age of female geriatric infirmity (yr) | $\min\{2L, 75\}$ |
| $\rho_1(L)$ | Age male eligibility starts (yr) | $\min\{L, 15\}$ |
| $\rho_g(L)$ | Age male eligibility ends (yr) | $\min\{2L, 75\}$ |
| $c(M)$ | Rate of female conception and delivery (1/yr) | Decreasing function of $M$: $c(M) = 3e^{-0.1(M-45)}$ |
| $\alpha(L)$ | Male weighting factor for paternity competition | Decreasing function of $L$: $\alpha(L) = \exp\left\{-0.36\int_0^L e^{-0.07u}\,du\right\}$ |
| $\beta$ | Reduction of dependent mortality from grandmother | 0.9 |
| $C_{max}$ | Max. subsidy cost for a just-weaned dependent | 0.5 |
| $D_{max}$ | Max. number of dependents a grandmother can subsidize at once | 6 |
| $p$ | Probability of mutation in $L$ and $M$ at birth | 2% |
| $\sigma$ | Standard deviation of mutations | 2% |
| $K$ | Population carrying capacity | 1,000 |

*Parameter estimates*

Since a mother who is pregnant is not eligible for another conception, we include gestation in an individual's age, making it time since conception rather than since birth. We call the age at which an infant could survive without its mother the age of weaning, and estimate it to be $\tau_0(L) = \min\{L/8, 3\}$. For example, if $L = 20$ or higher, which we take to represent the range of expected adult lifespans of great apes (17, 42), (but see (43)), then $\tau_0 = 2.5$. Since the gestation period of living hominids, humans and great apes is approximately 8.4 months, or 0.7 years (44-48), $\tau_0 = 2.5$ corresponds to an age of 1.8 years since birth. For $L = 24$ or higher, $\tau_0 = 3$, which corresponds to a weaning age of 2.3 years since birth. This estimate of the period of unweaned dependency is consistent with data from living people and other great apes (49, p. 50).

We set other life history parameters as follows: offspring become independent at age $\tau_1(L) = L/5.2$; females reach sexual maturity at age $\tau_2(L) = L/2$; female fertility ends at age $M$; and females reach infirmity and cannot grandmother at age $\tau_g(L) = \min\{2L, 75\}$. Based on reported age ranges of men's fertilities from ethnographers and reported by Tuljapurkar et al. (50), we assume that males become eligible to compete for paternities at age $\rho_1(L) = \min\{L, 15\}$; and males reach infirmity and cannot compete at age $\rho_g(L) = \min\{2L, 75\}$, just like the females, so that no paternities go to males beyond age 75.

Evidence from human populations practicing natural fertility (45, 51) shows, and observations of wild chimpanzees suggest (52), that time to next birth increases with parity, which suggests a monotonic tradeoff: higher $M$ gives an opportunity for more births, but also increases time to conception. Thus, we assume that the female conception rate, $c(M)$, is a decreasing function of the age female fertility ends, $M$. We chose the form of a decreasing exponential. For our system, the function $c(M) = 3\exp(-0.1(M - 45))$ produced results that fit the empirical pattern. Although the number of months to conception observed in great apes varies widely (46), human data (53) suggest an average of approximately 4 months. For our choice of parameters, when $M$ = 45 years, which is in the human and great ape-like range, the conception rate $c(45)$ is 3/yr, corresponding to an average time to conception of 1/3 year, or 4 months. On the other hand, if $M$ drops to 38 years, the conception rate $c(38)$ = 6/yr, corresponding to an average time to conception of 2 months. If $M$ rises to 52 years, the conception rate $c(52)$ = 1.5/yr, corresponding to an average time to conception of 8 months. Figure 3a shows a plot of conception rate, $c(M)$, versus age female fertility ends, $M$.

We assume that the parameters pertaining to the cost of juvenile subsidies are $C_{max}$ = 0.5 and $D_{max}$ = 6, which means each grandmother can support two just-weaned dependents up to a maximum of six dependents at once. In a later sensitivity analysis, we weaken the ability of grandmothers to subsidize dependents and investigate resulting outcomes.

We determine the male weighting factor $\alpha(L)$ in terms of its proportional change $\delta(L) = -\alpha'(L)/\alpha(L)$. Since we assume $\alpha(L)$ is a decreasing function, the function $\delta(L)$ is nonnegative, and $\delta(L)$ uniquely determines $\alpha(L)$ up to a scaling factor by the expression

$$\alpha(L) = \exp\left\{-\int_0^L \delta(u)\,du\right\}.$$

We assume that $\delta(L)$ is of the form $\delta(L) = \delta_0 \exp(-\rho M)$. In other words, as $L$ increases, the relative change in male weighting decreases exponentially. For our system, the function $\delta(L) = 0.36e^{-0.07L}$ gave rise to equilibria at chimpanzee-like and human-like values of life expectancy. Figure 3b shows a plot of the male weighting function, $a(L)$, versus the expected adult lifespan, $L$.

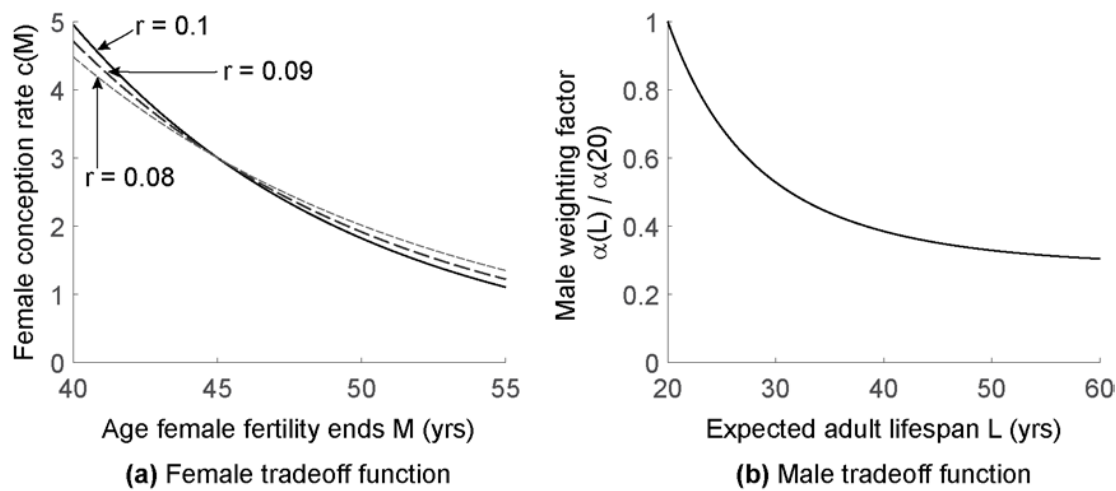

**(a)** Female tradeoff function          **(b)** Male tradeoff function

**Figure 3**. (a) Female conception rate $c(M) = 3\exp\{r(M - 45)\}$ for three different tradeoff intensities $r = 0.1$, $0.09$, and $0.08$. These three cases are compared in the *Sensitivity analysis* subsection of *Results*. (b) Male weighting factor $\alpha(L) = \exp\left\{-\int_0^L 0.36e^{-0.07u}du\right\}$, normalized by $\alpha(20)$ to make the curve scale to 1 at $L = 20$.

For the reduction in mortality of dependents from having a living grandmother, we set $\beta = 0.9$, or $90\%$. This value means that a dependent with a longevity ranging from $L = 20$ to $40$ with a living grandmother has a 25% to 13% increased chance of surviving to age 5, (calculated as $\exp(5(1 - \beta)/L)/(\exp(5/L) - 1)$. (Blurton Jones (39: Table 18.2) estimates a 65% increased chance of survival from his Hadza data, but this level of increase is not attainable for the model at hand even if we reduce dependent mortality to 0, because of our assumption of constant lifelong mortality of $1/L$.) For completeness in our sensitivity analysis, we consider reduced values of $\beta$ to investigate the influence of this parameter.

We assume that the probability $p$ of mutation per conception is 0.02 as in Coxworth et al. (54), lower than in Kachel et al. (55) and Kim et al. (35) which used 0.05, and we assume that mutational shifts in $L$ and $M$ have a standard deviation of $\sigma = 0.02$, or 2%. These values do not make a difference except on the rate of transition from a great ape-like to human-like equilibrium.

We assume that the population carrying capacity $K = 1,000$. This value allows the population to possess sufficient size and heterogeneity to produce consistent results. On the other hand, it is small enough to allow simulations to run quickly. Table 1 lists parameters and estimated values.

We begin each simulation with 500 females and 500 males with expected adult lifespans of $L = 20$ (representing a great ape-like expected adult lifespan) and ages $15 + 20n/365$, for $n = 0, \ldots, 500$, i.e., the female and male populations consist of individuals with ages spaced 20 days apart. We apply this initial state simply as a seed population. The system rapidly converges to a steady great ape-like age and demographic distribution within several generations, which is negligible in the evolutionary timescale of the model.

*Methods for sensitivity analysis*

We use the parameter values listed in Table 1 as starting estimates, but to investigate sensitivity, we conduct additional simulations under varying parameters related to the female fertility tradeoff and the strength of grandmother help. In particular, we vary $C_{max}$, the maximum cost of the subsidy required for a just-weaned dependent; $D_{max}$, the maximum number of dependents one grandmother can adopt at once; $\beta$, the reduction of dependent mortality from grandmother benefit; and the tradeoff of the female fertility function by varying the exponent $r$ of $c(M) = 3e^{-r(M-45)}$.

In addition, as mentioned in the *Grandmothering* section, for our default algorithm, eligible grandmothers adopt matrilineal grandchildren from eldest to youngest, followed by matrilateral nieces and nephews from eldest to youngest. To investigate the model's sensitivity of this assumption, we also consider the following alternatives: (no aunting) grandmothers only adopt matrilineal grandchildren, and (youngest adopted first) grandmothers adopt youngest matrilineal grandchildren first and the youngest matrilateral nieces and nephews first.

**Results**

We first simulate the system with and without grandmothering. Parameter values are as in Table 1. Results of 50 simulations without grandmothering up to a time of 1 million years are shown in Figure 4a.

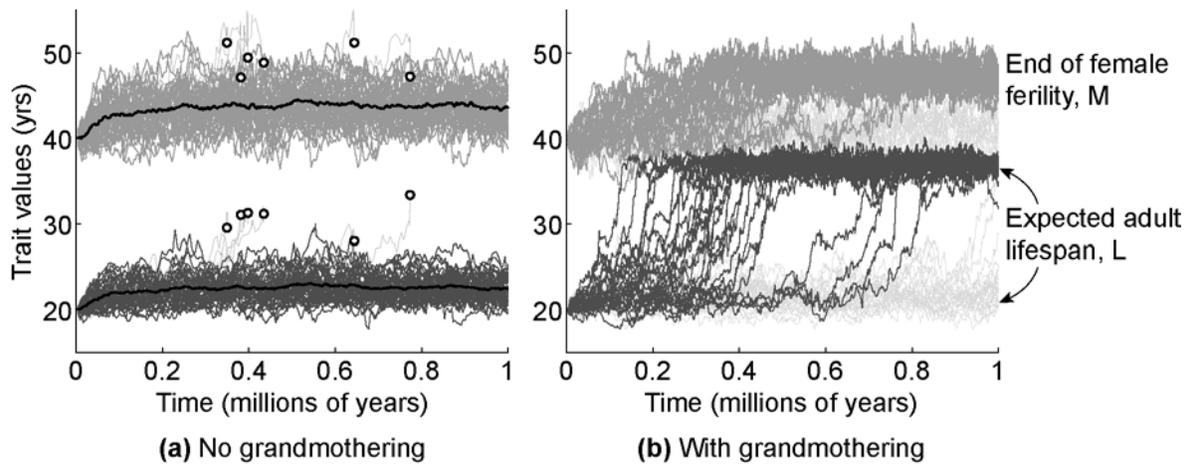

**(a)** No grandmothering   **(b)** With grandmothering

**Figure 4.** (a) Evolution of population-average expected adult lifespan, *L*, (darker gray) and end of female fertility, *M*, (lighter gray) for 50 simulations in the absence of grandmothering. Six simulations, terminated with white circles, resulted in extinction before 1 million years. Average *L* and *M* for populations that did not go extinct are shown in black. (b) Evolution of population-average expected adult lifespan, *L*, and end of female fertility, *M*, for 50 simulations in the presence of grandmothering. Twenty-six simulations (indicated by the darker shades of gray) shift out of the region of great ape-like longevity toward human-like longevity. The other 24 simulations are shown in very light gray. For both cases, parameters are taken from Table 1.

In Figure 4a, we see that 44 of 50 simulations survived up to 1 million years. The other 6 simulations went extinct when the population-average expected adult lifespan, *L*, rose to between 28 to 32 years and the population was no longer viable. Of the 44 simulations that survived, the overall average *L* from 200,000 to 1 million years ranged from 22.0 to 23.1 years and the overall average *M* over the same time period ranged from 43.0 to 44.6 years of age. These values of *L* and *M* represent a great ape-like equilibrium, and our results show that a large majority of simulations can survive up to at least 1 million years, indicating that the population without grandmothering shows long-term stability at these values of *L* and *M*.

In Figure 4b, we see that within 1 million years, 26 of 50 simulations shift from the great ape-like to the human-like equilibrium of increased longevity. The other 24 simulations remain at the great ape-like equilibrium or only begin to exit this equilibrium after 900,000 years. These results indicate that the great ape-like equilibrium is a locally stable fitness peak and that this peak exists at the brink of a transition to a higher fitness peak, corresponding to the human-like equilibrium. The very light gray curves that do not shift to human-like longevity within 1 million years will get there later. In fact, if we consider escape from the great ape-like equilibrium to be the time when the average population longevity trait exceeds 33 years, then the average period of escape for these simulations is 82,000 years, which means we expect that with grandmothering subsidies all 50 populations will escape to human-like longevity after approximately 50 x 82,000 = 4.1 million years.

In Figure 4a, we see that 6 of the 50 simulations result in extinction after the expected adult lifespan exceeds 25 years. Investigating this phenomenon further, we can show that the human-like equilibrium of higher longevity is not viable without grandmothering, because the net growth rate falls below 0; on the other hand, with grandmothering subsidies, the higher equilibrium longevity becomes not only viable but likely.

The other key observation is that even as longevity in the population increases to human-like values in the presence of grandmothering, the end of female fertility does not increase comparably. In particular, for the populations that did not go extinct in Figure 4a, the population-average expected adult lifespans, $L$, at the end of the 1 million years is 22.4, and the average age female fertility ends, $M$, is 43.7. On the other hand, for the populations that pass $L =$

30 by 900,000 years in Figure 4b, the average *L* and *M* at the end of 1 million years are 36.1 and

46.5, respectively, so we see that a 65% increase in population-average expected adult lifespan

only results in an 8.7% increase in the age female fertility ends during the transition from great

ape-like to human-like equilibria.  For comparison, if we remove aunting and assume only

grandmothering the average *L* and *M* at the end of 1 million years hardly changes to 36.7 and

46.6, respectively (data not shown).

*Sensitivity analysis*

As described in *Methods for sensitivity analysis*, we investigated several parameters pertaining to female fertility and grandmother help and their impact on the long-term evolution of the population.  Table 2 summarizes outcomes of 50 simulations under the various parameter changes from estimated parameters in Table 1.

**Table 2.** Outcomes of 50 simulations under various parameter changes from the base parameters from Table 1. Outcomes without aunting, i.e., matrilineal grandmothering only, are shown in parentheses. Notes: The labelled cases (a) resulted in one extinction, (b) resulted in 4 extinctions, and (c) resulted in 10 extinctions.

| Varied parameter | Grandmothering + aunting | Grandmothering, no aunting |
|---|---|---|
| Table 1 parameters | 52% increased longevity | 48% increased longevity |
| Youngest adopted first | 60% | 46% |
| Care cost $C_{max} = 1$ | 38% | 26% |
| Max dependents $D_{max} = 2$ | 52% | 54% |
| Max dependents $D_{max} = 1$ | 38%    (a) | 26% |
| Max dependents $D_{max} = 0$ | 0    (b) | 0    (c) |
| Mortality benefit $\beta = 0.5$ | 48% | 38% |
| Mortality benefit $\beta = 0$ | 26% | 14% |
| Fertility tradeoff $c(M) = 3e^{-r(M-45)}$, $r = 0.05$ to $0.15$ | Results in Figure 5 | Results in Figure 5 |

We divided our sensitivity analysis into four parts. First, we see that even if we double the cost of caring for dependents from $C_{max} = 0.5$ to $C_{max} = 1$, the population with grandmothering still shifts to increased longevity at a relatively high probability, so we conclude that grandmothering can still drive the evolution of increased longevity even if the value of her subsidy is reduced by half.

Next, we see that even after reducing the maximum number of dependents that one grandmother can subsidize to $D_{max} = 2$ and even $D_{max} = 1$, the population still shifts to increased longevity. In contrast, if we eliminate the ability of grandmothers to subsidize dependents by setting $D_{max} = 0$, we also eliminate the evolution of increased longevity. These results show that the ability of grandmothers to reduce interbirth intervals by subsidizing dependents is crucial for driving the evolution of increased longevity. Furthermore, even the ability to support only one dependent per grandmother can still reliably propel the evolution of longer lifespans. In addition, we see that reducing the benefit of grandmothering on lowering dependent mortality from $\beta = 90\%$ to 50% and even to 0 reduces the chance of a population transitioning to increased longevity, but it does not eliminate the transition altogether.

From our sensitivity analysis of these three parameters pertaining to the strength of grandmother help, we conclude that all parameters contribute to the chance of a population shifting to increased longevity, but subsidizing dependents and thereby reducing interbirth intervals is the crucial type of grandmothering necessary to guarantee that such a transition is possible.

To study the evolution of the end of female fertility, as a trait distinct from $L$ yet interacting together, we varied the female fertility tradeoff $c(M) = 3e^{-r(M-45)}$ by varying the exponent $r$ from 0.05 to 0.15 and recording the long-term population averages of $L$ and $M$ with and without grandmothering. The results of 50 simulations for each value of $r$ are shown in Figure 5. As expected, increasing (or decreasing) the cost of extending fertility to older ages shifts $M$ to lower (or higher) ages both with and without grandmothering. On the other hand, one striking result is that grandmothering hardly shifts the long-term average $M$ even as the long-term average $L$ increases to higher ages (see Figure 5). These results show that even with varying female fertility tradeoffs, grandmothering more strongly influences the evolution of increased expected adult lifespan and only weakly influences the evolution of the end of female fertility, if at all.

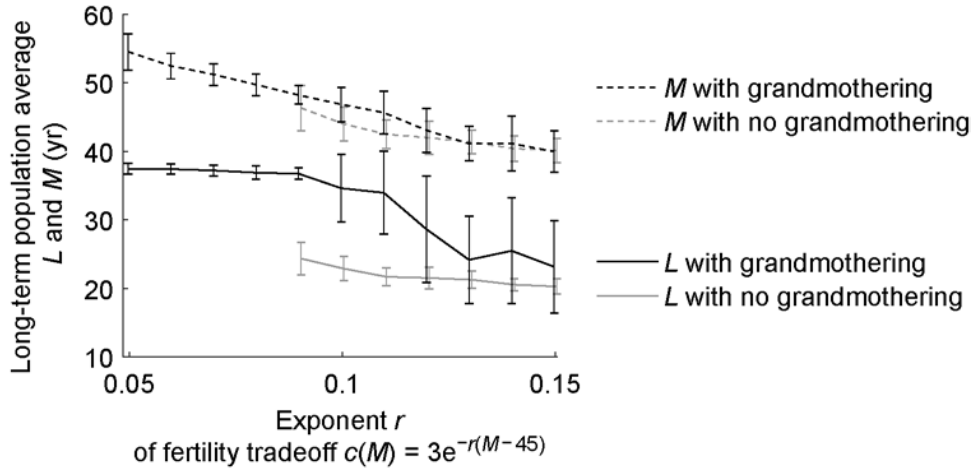

**Figure 5.** Long-term population averages of 50 simulations showing expected adult lifespan ($L$) and end of female fertility ($M$) for varying exponents $r$ of the female fertility tradeoff $c(M) = 3e^{-r(M-45)}$. (Higher $r$ corresponds to higher costs of increasing fertility to older ages.) Cases with grandmothering and with no grandmothering are considered, and error bars indicate standard deviations. All simulated populations with no grandmothering go extinct within 1 million years when $r \leq 0.08$, so no results for $L$ and $M$ are shown.

We note that as $r$ decreases lower than 0.09, the populations without grandmothering go extinct within 1 million years. In fact, even at $r = 0.09$, the end of female fertility begins to increase, which leads longevity to increase, which in turn influences $M$ to increase further, which in turn favors increased longevity until – without grandmothering – 21 out of 50 of these populations reach extinction as $L$ exceeds around 26. These results show that a sufficiently strong cost to extending fertility, e.g., a female fertility tradeoff $c(M)$ as suggested empirically (52), is required for the great ape-like equilibrium to be stable and viable in the absence of grandmothering. For all values of r where the populations are viable without grandmothering (i.e., $r \geq 0.09$), the average end of female fertility with and without grandmothering remains around or below 50 years as observed for humans and great apes.

Finally, we consider the effect of the assumption that post-fertile aunts also contribute to provisioning matrilateral nieces and nephews. We remove aunting from all scenarios. In all cases, (except possibly the case where the maximum number of dependents that can be subsidized $D_{max} = 2$), aunting increases the chances of shifting to higher longevity, but the qualitative results are the same with and without aunting (see Table 2). We note that this outlier case for $D_{max} = 2$ is likely due to probabilistic noise. A more thorough analysis with higher numbers of simulations would be necessary to statistically determine the significance of aunting in all cases.

**Discussion**

Our agent-based model investigates the coevolution of expected adult lifespan and end of female fertility. We allow both of these traits to evolve in the population and find that grandmother effects can drive the population from an equilibrium representing a great ape-like longevity to a human-like longevity, while maintaining the observed termination of women's fertility before the age of 50.

Simulations reveal complex dynamical interactions showing that, as classic evolutionary theories of aging assume (4, 30, 31, 56), likely survival and future fertility are highly coupled. In most circumstances, increasing somatic longevity will drive a comparable increase in the end of female fertility, and conversely, an increase in the age female fertility ends will drive an increase in somatic longevity to raise benefits in lifetime reproductive success. However, as Fisher (30) recognized, selection maximizes reproductive value, which includes not only offspring but ancestry of future generations. As in his explanation for offspring sex ratios, the fitness payoff for sons (vs daughters) must be measured in grandchildren (30, 57). In a similar way, increased longevity in the modeling here pays off not in future fertility but increased numbers of heritable variants one generation removed.

In our investigation, we find that to maintain a great ape-like life history of lower longevity and comparable end of female fertility, both traits need tradeoffs that prevent the traits from coevolving to unrealistically high levels beyond a great ape-like equilibrium. As found in Kim et al. (35), we also see that a great-ape like longevity is maintained by fertility-longevity tradeoffs for both females and males. However, in this work, we discover the additional need for a female tradeoff between conception rate and end of fertility. As we need a tradeoff for longevity versus fertility to limit cost-free evolution toward greater and greater longevity, we also need a tradeoff involving the end of female fertility to prevent cost-free evolution toward greater and greater end of female fertility.

After obtaining a great ape-like equilibrium, our simulations show that the grandmother effect primarily shifts the longevity tradeoff, leading to an increased human-like longevity; however, the grandmother effect does not substantially increase the end of female fertility. As a result, we arrive at a human-like life history equilibrium of increased longevity without an increased end of female fertility. As a simplification, we assume grandmothering only occurs after the end of female fertility at age $M$, imposing a tradeoff between continued fertility and grandmothering, which could be further investigated in future modelling. As in our previous study, we also find that grandmothering has its strongest effect from decrease interbirth intervals rather than reducing offspring mortality (35).

As a future continuation of our work, a broad and open question is that if $L$ and $M$ can both freely mutate, why does primate post-fertile longevity only appear in the hominid radiation, rather than also at monkey-like life histories, e.g., why don't at least some of the macaques or baboons that reach menopause at age 25, also have substantial post-fertile life stages? The only other mammalian radiation in which post-fertile life stages evolved, the toothed whales, includes very long-lived taxa (58, 59). Perhaps a longevity threshold is a necessary, but (as elephants and some cetaceans show) not sufficient, condition for the evolution of post-fertile longevity. The grandmother hypothesis for the evolution of human life history proposes a key role for ecology, ancestral populations facing habitats where handling requirements for important foods were beyond the capacities of just-weaned juveniles (24-27). Although we have restricted our inquiry to hominid parameters, our investigation is a framework for a mathematical modeling approach to begin asking whether interaction is required between longevity and ecology for a post-fertile stage to evolve. If grandmothering has been advantageous enough to propel the evolution of human post-menopausal longevity with its many consequences in our lineage (3, 54), why is it so rare (60, 61)? For future work, we propose to address these next questions by extending our current model beyond hominid parameters to include other primates and to couple the evolution of longevity with ecological changes that would have led to different availabilities of resources, resulting in differential impacts on infants, juveniles, and adults.

## Acknowledgements

We thank Nick Blurton Jones for valuable criticism and advice.  PSK was supported by the

Australian Research Council, Discovery Project (DP160101597).